# Evidence of zero-field Wigner solids in ultra-thin films of cadmium arsenide

**Simon Munyan, Sina Ahadi, Binghao Guo, Arman Rashidi and Susanne Stemmer***

Materials Department, University of California, Santa Barbara, CA 93106-5050, USA.

*Email: stemmer@mrl.ucsb.edu

**Abstract**

The quantum Wigner crystal is a many-body state where Coulombic repulsion quenches the kinetic energy of electrons, causing them to crystallize into a lattice. Experimental realization of a quantum Wigner crystal at zero magnetic field has been a long-sought goal. Here, we report on the experimental evidence of a Wigner solid in ultra-thin films of cadmium arsenide ($Cd_3As_2$) at zero magnetic field. We show that a finite bias depins the domains and produces an unusually sharp threshold current-voltage behavior. Hysteresis and voltage fluctuations point to domain motion across the pinning potential and disappear at finite temperature as thermal fluctuations overcome the potential. The application of a small magnetic field destroys the Wigner solid, pointing to an unconventional origin. We use Landau level spectroscopy to show that the formation of the Wigner solid is closely connected to a topological transition as the film thickness is reduced.

# I. INTRODUCTION

Wigner crystals in solid-state systems are a highly sought-after, many-body phase that arises from electron-electron interactions and was first theoretically predicted nearly a century ago [1]. A Wigner crystal is realized when Coulombic repulsion between electrons dominates over the kinetic energy and the electrons crystallize into a lattice. To date, Wigner solids have been observed at high magnetic field [2-5], which serves to suppress the kinetic energy, and in two-dimensional van der Waals materials, where the Wigner solid is stabilized by a moiré lattice potential [6,7]. Only a few reports exist of Wigner solids forming spontaneously at zero magnetic field [8-11]. An alternative route to a Wigner solid involves strongly spin-orbit coupled two-dimensional electron gases. In such systems it has been predicted that even short-range interactions can produce Wigner phases, which, moreover, may also host rich magnetic and other symmetry-breaking ground states [12-14]. Remarkably, this route has not been explored experimentally.

Here, we report on experimental signatures of hole and electron Wigner solids in dilute, ultra-thin films of cadmium arsenide ($Cd_3As_2$), a strongly spin-orbit coupled topological material. These Wigner solids feature distinct and remarkably clear transport signatures uniquely associated with depinning the solids from the disorder potential. These include threshold and hysteretic current-voltage (I-V) behavior and sawtooth voltage fluctuations below the depinning threshold voltage. The latter points to theoretically predicted stick-slip motion of domains following a ratchet mechanism, which has not been observed previously. All of these signatures disappear simultaneously at a critical temperature, which we attribute to depinning of the Wigner solid. Surprisingly, the Wigner solid is destroyed (rather than stabilized) by a magnetic field, which is suggestive of an unconventional origin. We discuss the results in the context of the band structure of a topological insulator in the presence of Rashba spin-orbit coupling, which can explain the origin of the Wigner solid.

## II. EXPERIMENTAL

Epitaxial thin films of (001) $Cd_3As_2$ were grown on an $Al_{0.45}In_{0.55}Sb$ buffer layer on a (001)-oriented GaSb substrate via molecular beam epitaxy, as reported elsewhere [15]. The $Cd_3As_2$ film thicknesses were 16 nm and 13 nm, respectively. Gated Hall bar devices with a mesa width of 50 µm were patterned using standard photolithography and lift-off techniques. Device mesas were defined with argon ion milling. Ohmic contacts were deposited via electron-beam deposition of Ti/Pt/Au (20/20/200 nm). Devices were treated with a low-power oxygen plasma ash prior to deposition of a blanket $Al_2O_3$ gate dielectric via atomic layer deposition at 120 °C. The dielectric was etched from the ohmic pads with dilute tetramethylammonium hydroxide. Lastly, a gate electrode of Ni/Au (5/180 nm) was deposited via thermal evaporation. All magnetotransport measurements were performed in a He-3/He-4 dry dilution refrigerator with a base temperature of 12 mK unless stated otherwise. AC resistance measurements were performed using lock-in amplifiers under current bias at 17.778 Hz (see Fig. S1 [16]). All other measurements were performed at DC. For the DC measurements, the longitudinal voltage across the device was measured using a voltage preamplifier connected to a voltmeter. Current was measured using a current preamplifier and read out by a voltmeter. DC voltage sources supplied the source-drain bias and gate bias. Additional information can be found in the Supplementary Notes [16].

## III. RESULTS AND DISCUSSION

Figure 1a shows the Landau level spectrum of a 16 nm (001) $Cd_3As_2$ film under an applied out-of-plane magnetic field and gate voltage at 12 mK. The insulating gap at charge neutrality is indicated by the blue star. Unlike slightly thicker films (see ref. [17]), which show a crossing of the zeroth Landau levels that is characteristic of the inverted band structure of a two-dimensional topological insulator, this film shows an *avoided* crossing of the zeroth Landau levels. In addition to the charge neutrality gap, another region of low conductivity appears in the valence subband (red square) at a carrier density of $8.8 \times 10^{10}$ cm$^{-2}$ (see Fig. S2 [16]), which persists up to moderate magnetic fields until it is interrupted by the emergence of a $\nu = -1$ quantum Hall state. At zero field this region is visible as a peak in resistance, which is distinct from the smaller peak corresponding to charge neutrality (Fig. 1b). In the following, we will show experimental signatures that suggest that this resistive region, which is not present in slightly thicker films that are in the topological insulator phase (see ref. [17]), signifies a hole-like Wigner solid. The local minimum between the two peaks corresponds to the subband from which the hole-like zeroth Landau level originates. The 13 nm film (Fig. 1c) shows a wider and more resistive gap at charge neutrality (see Fig. S3 [16]). The conventional dispersion of the zeroth Landau levels signifies a trivial insulator. The hole-like Wigner solid is also observed in this film (see green square in Fig. 1c), where it conspicuously interrupts the hole-like zeroth Landau level. In addition, this film also has an electron-like Wigner solid near the conduction subband edge in the charge neutrality gap (orange diamond), as discussed below.

The first signature supporting the Wigner solids is the unique double-threshold and hysteretic behavior of I-V traces when the Fermi level is tuned into one of the new resistive states (see Fig. 2 for the 16 nm film). The double-threshold behavior is characterized by a sharp increase in current, followed by a linear region as bias is swept in the forward direction. The dynamic threshold ($V_d$) occurs at the bias intercept of the linear regime. The static threshold ($V_s$) is the point where current sharply increases. In the reverse sweep direction, the I-V forms a hysteresis loop where the linear behavior is retained below $V_s$ until it sharply drops to zero at the retrapping threshold ($V_r$). Both double-threshold behavior and hysteresis are unique to the Wigner solid phases in both films, which can be seen by examining the I-V traces across a range of gate voltages (Fig. 3). In the 16 nm film (Fig. 3a), the traces just below charge neutrality show the double-threshold I-V behavior, when the Fermi level is tuned into the hole-like Wigner solid state. The I-V traces are ohmic far above and below charge neutrality in the conduction and valence subbands, respectively. The same behavior appears in the hole-like Wigner solid state of the 13 nm film (Fig. 3b). The traces spanned by the black arrow were taken in the charge neutrality gap, which is exceedingly resistive (> 100 MΩ) in this film [see Figs. S3 and S4 for raw data] [16]). Notably, the traces just above the gap are also non-linear. A finer scan reveals the same characteristic I-V behavior, indicating the presence of a second, electron-like Wigner solid (see Fig. S5 [16]). Additionally, there is a small diode effect between the positive and negative static thresholds in both films, which will be discussed later.

The threshold behavior and hysteretic I-V are distinct signatures of a pinned Wigner solid [9,18-21]. For $V_{bias} < V_s$, the Wigner solid is pinned by the disorder potential and no measurable current flows (< 1 pA). At $V_{bias} = V_s$, the electric field depins the Wigner solid, resulting in a sharp increase in current. For $V_{bias} > V_s$, the Wigner solid slides, resulting in a linear I-V. The hysteresis can be explained in models that consider domain motion [22].

Before turning to additional characteristics of the Wigner solids, we briefly rule out potential alternative explanations for the observed I-V. Source and drain contacts may influence

I-V curves in a two-probe measurement and produce spurious non-linearity, particularly at low carrier densities. However, the I-V is linear and non-hysteretic in the $\nu = -1$ quantum Hall plateau at the same carrier density as the Wigner solid (Fig. S6 [16]). The contact resistance determined from this state is negligible (7 kΩ) compared to the differential resistance of the pinned WS (~100 MΩ). More trivial explanations such as percolation and single-particle localization cannot explain the double-threshold behavior [23]. Similar behavior has also appeared in pinned charge-density waves (CDWs). For CDWs, however, the threshold voltage diverges at the Peierls temperature [24-26]. In contrast, we observe a monotonic decrease in the threshold voltage with increasing temperature up to the critical temperature (Fig. 4a), which rules out the conventional CDW as an origin, but is consistent with the behavior of Wigner solids [19].

We can also rule out electron overheating at very low temperatures as the origin of non-linear I-V [27,28]. Power dissipation in our devices is $< 10^{-14}$ W below $V_s$ which is negligibly small. However, above $V_s$ current suddenly increases and the power rises to $\sim 10^{-12}$ W. If overheating is significant above $V_s$, then the hysteresis should be sensitive to the bias sweep rate. However, we find no correlation between them under repeated bias sweeps (Fig. S7 [16]). In the case of overheating, current increases exponentially above the threshold as the electron temperature decouples from the phonon bath and rapidly increases [28]. The double-threshold behavior we observe has a distinctly linear behavior above $V_s$, in contrast to previous reports of current jumps in disordered superconductors [27,29]. Additionally, the differential resistance at zero bias does not saturate from 100 mK down to 15 mK, indicating that the electrons remain coupled to the phonon bath for small biases (Fig. S8 [16]). Lastly, we note that in both samples, the distinct I-V characteristics are absent in the charge neutrality gap, both when it is more resistive (13 nm) and less resistive (16 nm) than the Wigner solid (Figs. 3a and 3b). Heating should be heightened in the 13 nm film at charge neutrality given its exceedingly large resistance (>100 MΩ). However, in this film the I-V traces are linear up to large biases well beyond $V_s$ (up to ~ 100 mV), where an exponential I-V appears due to thermionic emission over the charge neutrality gap (Fig. S4c [16]).

We next turn to the other characteristics of the Wigner solids, namely their temperature-dependent behavior and a unique voltage ratchet effect. The double-threshold and hysteresis disappear around the critical temperature $T_c$ = 65 mK (see Fig. 4a for the 13 nm film). The I-V remains non-linear above $T_c$ up to ~ 1 K (Fig. S9 [16]). Below $T_c$, the threshold voltage decreases with increasing temperature. This reduction in threshold and hysteresis reflects the increasing strength of thermal fluctuations, until the Wigner solid is depinned at $T_c$. The smooth, non-linear I-V above $T_c$ suggests a solid-liquid mixture [30,31]. The transition is also evident from a change in the activation energy at $T_c$ (Fig. S8 [16]). The temperature dependence of the 16 nm film is similar and $T_c \approx$ 65 mK (Fig. S10 [16]), which points to the intrinsic disorder of the material, expected to be similar in both films, as the origin of pinning.

Another remarkable feature of the Wigner solids are sawtooth voltage fluctuations below the critical temperature (Fig. 4b). These appear below $V_d$ and disappear abruptly at $V_s$, where the Wigner solid is depinned. Notably, the sawtooth pattern inverts at an intermediate bias voltage before disappearing. Like the threshold and hysteresis, the fluctuation amplitude also decreases with temperature and disappears at $T_c$ as the Wigner solid depins. The sawtooth shape is suggestive of stick-slip motion of domains, because they vanish at $V_s$ where the Wigner solid domains are depinned and because they appear only in the Wigner solid phase (Fig. S11 [16]). The fluctuations appear far below $V_d$, which may be expected for individual domain motion within the pinned phase [21]. We note that these fluctuations are different from *time-domain* fluctuations in CDWs [32,33] and circuit instability [20,34], as both are observed only above the threshold and at finite frequencies rather than DC [20,32]. Instead, the sawtooth shape we observe resembles a ratcheting mechanism predicted recently (ref. [35]). The Wigner crystal ratchet occurs due to asymmetry in the pinning potential, where domains move preferentially along the easy driving direction. The small diode effect which is generically observed in our Wigner solids (Fig. 3) also points to an asymmetric pinning potential [35], though the origin of the asymmetry requires further investigation. One possibility is the miscut of our substrate. Indeed, vortex ratchets have been theoretically studied in superconductors with patterned asymmetric thickness profiles [36].

Interestingly, ref. [35] predicts a crossover to a negative ratchet effect due to charge correlations, analogous to the inversion of the sawtooth pattern in our data (Fig. 4b). By reversing sweep direction, we also observe an inversion of the sawtooth pattern while preserving the crossover behavior, in good agreement with a ratcheting mechanism (Fig. S12 [16]).

Finally, we discuss the possible origins of the Wigner solids in $Cd_3As_2$ films. To this end, we first note that the appearance of the Wigner solid is unique to ultra-thin films of $Cd_3As_2$ near and below the topological transition. Near this transition, the Wigner solid emerges simultaneously with the appearance of an avoided crossing between the zeroth Landau levels (Fig. 1a). In contrast, thicker films (18 nm – 22 nm) feature a crossing of the zeroth Landau levels characteristic of a 2D topological insulator and do not show any unusual insulating states outside of the charge neutrality gap [17]. Although the transition to a trivial insulator is expected at low thicknesses from theoretical predictions [37,38], the appearance of an avoided crossing signals a transition driven by breaking of the $C_4$ rotational symmetry. This symmetry breaking would be consistent with the expected $C_6$ rotational symmetry of a Wigner crystal, though further investigation is needed. The emergence of Wigner solids in strongly spin-orbit coupled thin films suggests that the Rashba effect may play a role. The Rashba effect can arise due to spatial inversion asymmetry (SIA) of our heterostructure, which induces a perpendicular displacement field and spin splitting of the subbands (Fig. 1d). Crucially, the model shown in Fig. 1(d) results in van Hove singularities at the band edges, i.e., a sharp increase in the density of states and reduced kinetic energy, promoting the effects of electron correlations. In agreement with the model shown in Fig. 1(d), the Wigner solid symmetrically emerges near both band edges in the 13 nm film (Fig. S5 [16]). A weaker displacement field in the 16 nm film and heavier holes in $Cd_3As_2$ may be responsible for stabilizing the hole Wigner solid before its electron counterpart. Notably, the Rashba effect has been proposed to stabilize new types of Wigner solids even for short range interactions by creating a van Hove singularity at the band edges [12,13]. The presence of van Hove singularities, combined with the roles of short range interactions and spin-orbit coupling [12], makes conventional estimates of the stability range of the Wigner solid, based on the interaction parameter [39], difficult. Nevertheless, the sharp reduction of kinetic energy is clearly seen by the fact that the Wigner solid interrupts the hole-like zeroth Landau level in this film. An unconventional origin would also explain the very unusual behavior of the Wigner solid in a magnetic field. In both films, the double-threshold and hysteretic I-V disappear under a small out-of-plane magnetic field and eventually become linear at moderate fields (Fig. S13 [16]). The hole-type Wigner solid borders (and even possibly competes with) a $\nu = -1$ quantum Hall state. In the case of a more conventional Wigner solid, magnetic fields are expected to stabilize the Wigner solid and are even essential. In the case of Rashba spin-split bands, however, a sufficiently strong magnetic field changes spin textures and may eventually eliminate the van Hove singularity. We leave further investigation of these ideas to future work.

IV. CONCLUSIONS

In summary, we have shown experimental evidence supporting the formation of electron and hole Wigner solids in ultra-thin films of $Cd_3As_2$ that do not rely on magnetic fields for their stabilization. The Wigner solid appears in films that are sufficiently thin, presenting a new route to these long-sought zero-field electron crystals. The exquisite tunability of the system allows for accessing the physics of both electron and hole Wigner solids by simply tuning the gate voltage and displacement fields. The results offer new opportunities to understand the dynamics of pinned Wigner solids, including a unique voltage ratchet effect, which has been observed for the first time here. Perhaps the most exciting possibility is the variety of exotic ground states that strongly spin-orbit coupled Wigner phases have been predicted to host, which will be the focus of future studies.

**ACKNOWLEDGEMENTS**

The authors thank Xi Dai, Joel Moore, Andrea Young, Pavel Volkov, and Anton Burkov for helpful discussions. The work was supported by the Gordon and Betty Moore Foundation's EPiQS Initiative, Grant GBMF11944, and by a MURI program of the Air Force Office of Scientific Research under grant no. FA9550-22-1-0270. S.M acknowledges support from the Graduate Research Fellowship Program of the U.S. NSF (Grant No. 2139319) and the UCSB Quantum Foundry, which is funded via the Q-AMASE-i program of the U.S. NSF (Grant No. DMR-1906325).

**Figure Captions**

**Figure 1:** (a) Landau level spectrum of the 16 nm film at 12 mK. Filling factors obtained from the quantum Hall plateaus are labeled along with the pair of zeroth Landau levels. The blue star indicates the charge neutrality gap. The red square marks a gate voltage deep in the Wigner solid phase ($V_g$ = -2.08 V) which we use throughout this work to characterize the Wigner solid. (b) Zero-field longitudinal resistance of the 16 nm film at 12 mK. (c) Landau level spectrum of the 13 nm film at 2 K (see Supplementary Note 1 [16]). The magenta star marks charge neutrality, while the green square and orange diamond represent the hole-like and electron-like Wigner solids, respectively. (d) The effect of structure inversion asymmetry is a Rashba-like spin splitting, which creates van Hove singularities at the band edges (gray regions).

**Figure 2:** I-V taken in the Wigner solid ($V_g$ = -2.08 V, 12 mK) of the 16 nm film, showing double-threshold and hysteretic behavior. (Inset) Schematic of the Hall bar measurement configuration, where $V_{bias}$ is applied across the long arm and current is measured through the device, while a voltage $V_g$ is applied to the gate to modulate the carrier density.

**Figure 3:** I-V traces taken across a range of gate voltage in the (a) 16 nm and (b) 13 nm films at 12 mK. Traces are individually normalized to their maximum current and bias and offset by 0.5 to make qualitative differences between traces apparent. The traces spanned by the black arrow in (b) are not normalized, since the current remained below our noise threshold (1 pA). The gate voltage step between traces is $\Delta V_g$.

**Figure 4:** Temperature dependence of (a) I-$V_{bias}$ and (b) V-$V_{bias}$ from 15 to 100 mK in the 13 nm film. The temperature step between traces in both plots is $\Delta T = 5$ mK. Traces are offset by (a) 0.5 nA and (b) 2 mV, respectively. (Inset) Schematic of Hall bar measurement configuration. Here, V is measured between voltage probes along the longitudinal direction.

**Figure 1**

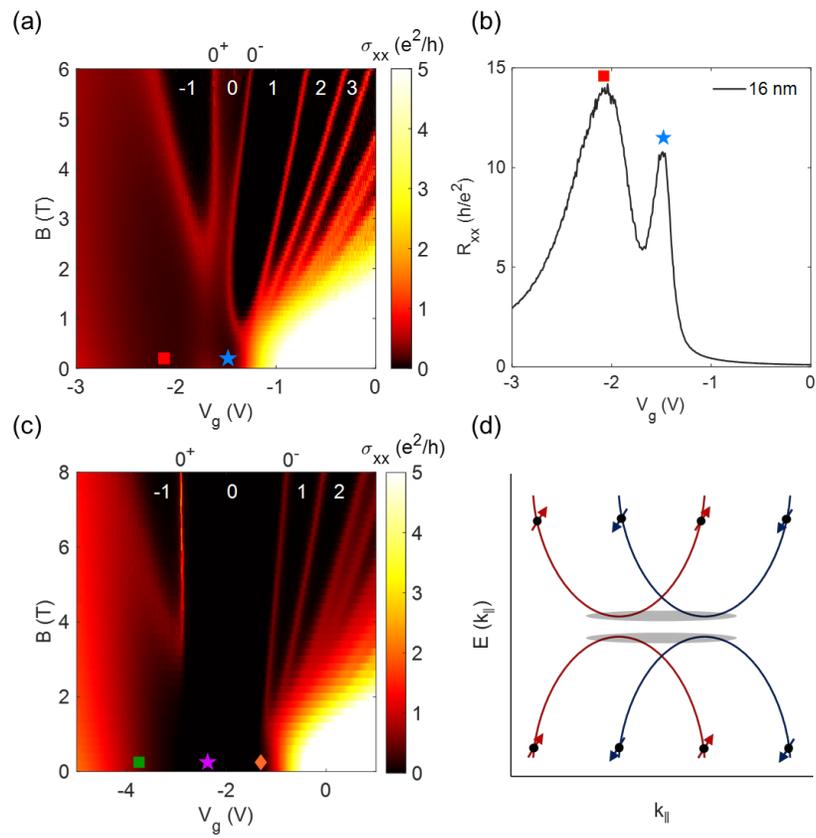

**Figure 2**

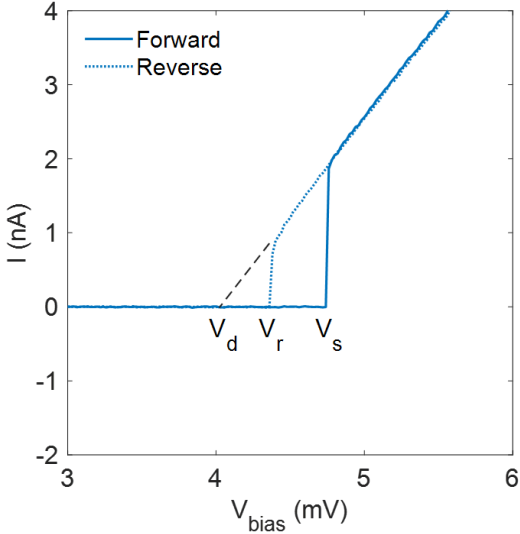

**Figure 3**

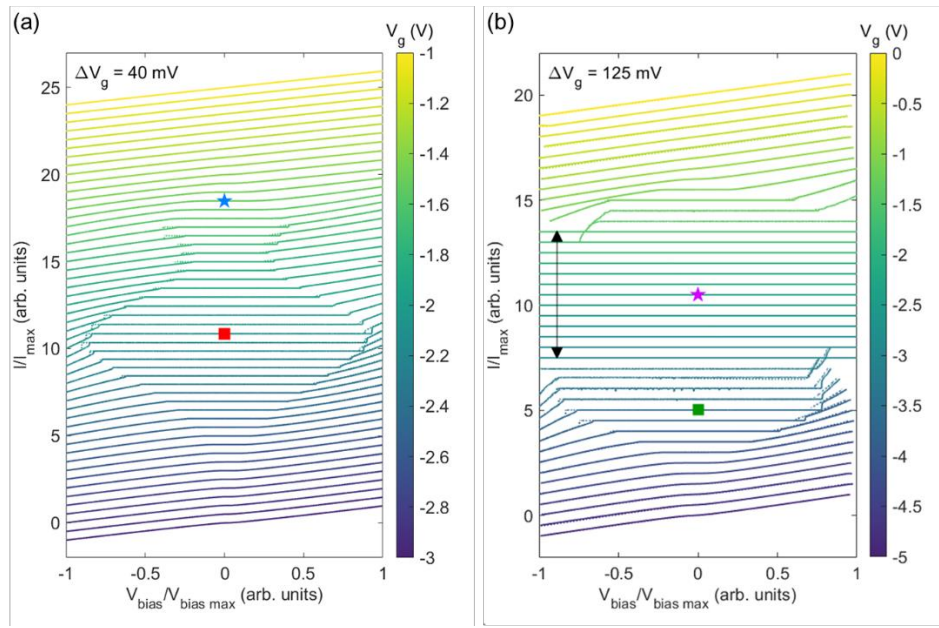

**Figure 4**

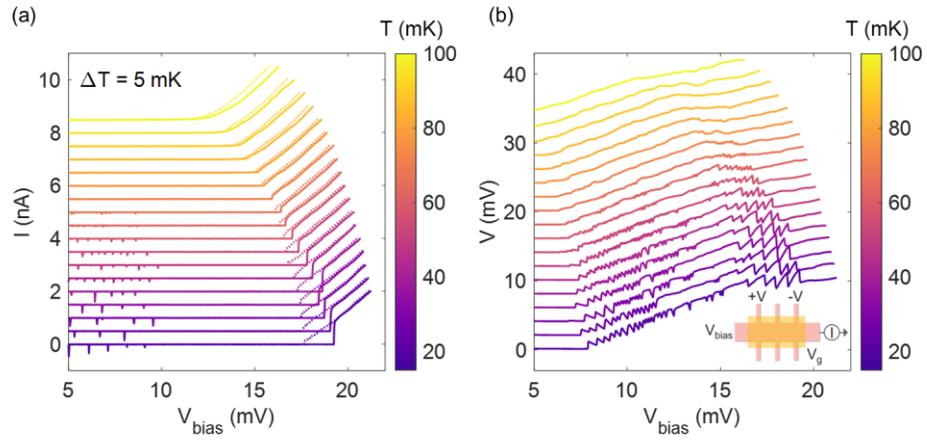

# Supplementary Information

# Zero-field Wigner solids in ultra-thin films of cadmium arsenide
**Simon Munyan, Sina Ahadi, Binghao Guo, Arman Rashidi and Susanne Stemmer[*]**

Materials Department, University of California, Santa Barbara, California 93106-5050, USA

[*]Email: stemmer@mrl.ucsb.edu

**Supplementary Figures**

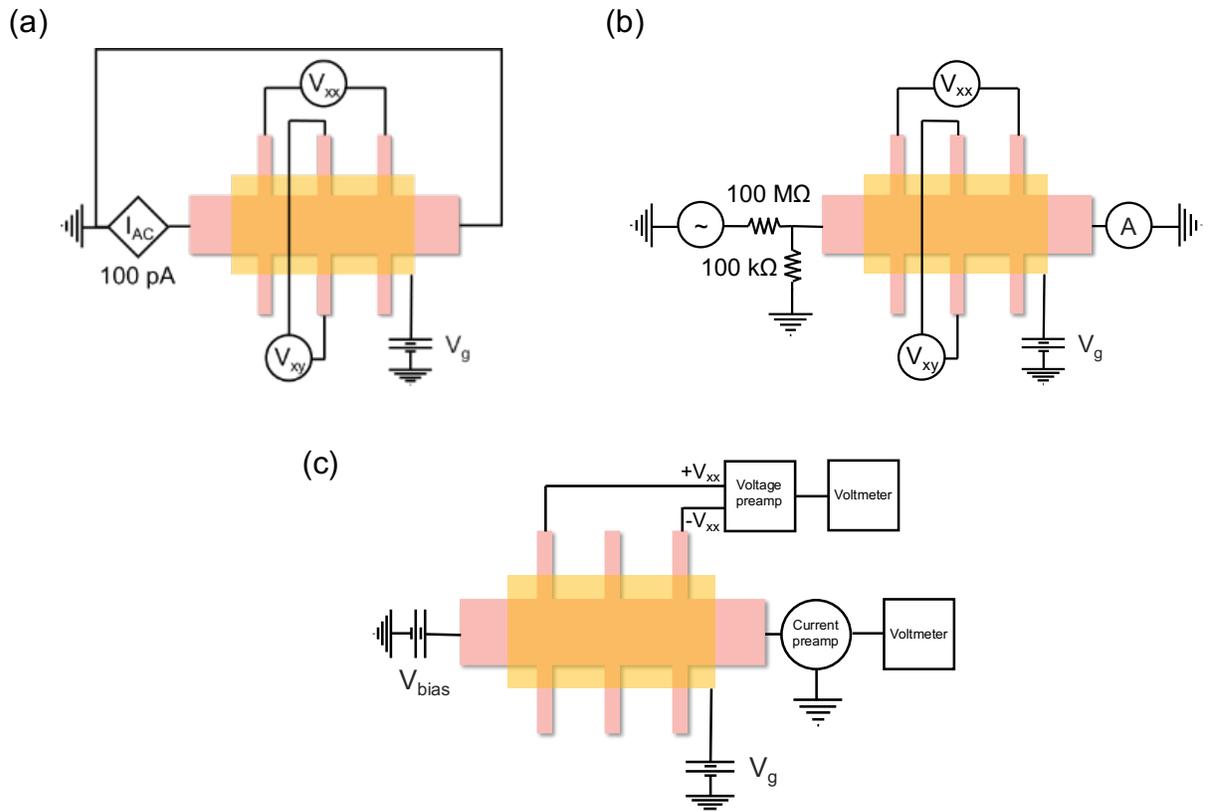

**Supplementary Figure 1**: Measurement circuits for the Landau level spectra of the (a) 16 nm film in Figs. 1a and 1b and of the (b) 13 nm film in Fig. 1c. (c) Measurement circuit for the DC I-$V_{bias}$ and V-$V_{bias}$ traces.

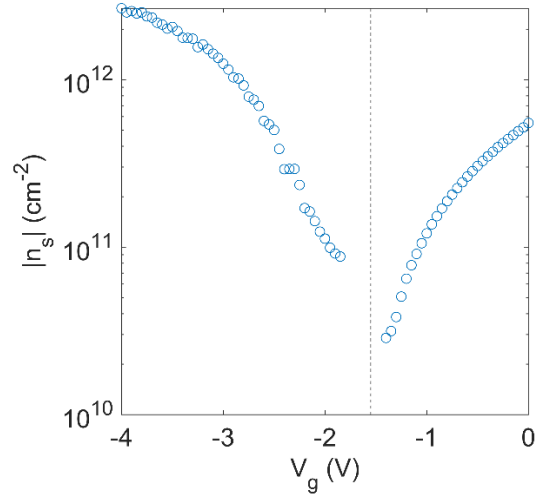

**Supplementary Figure 2:** Sheet carrier density ($n_s$) determined from Hall effect as a function of gate voltage ($V_g$) in the 16 nm film at 2 K. The vertical dashed line indicates charge neutrality. Carrier density could not be accurately determined near charge neutrality due to mixing of longitudinal and Hall channels.

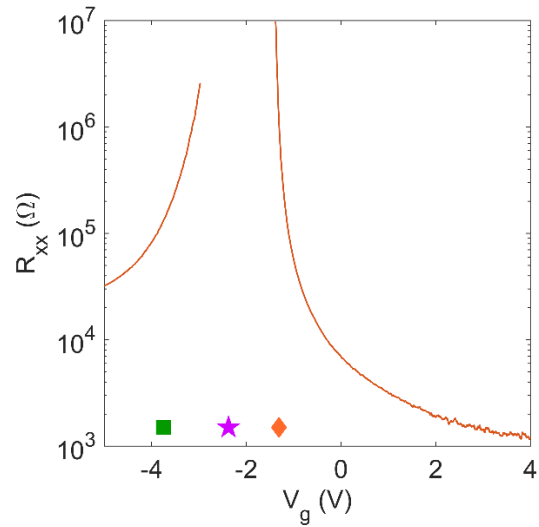

**Supplementary Figure 3:** Resistance of the 13 nm film as a function of gate voltage at 2 K. The green square, magenta star, and orange diamond mark the hole-type Wigner solid, charge neutrality point, and electron-type Wigner solid, respectively.

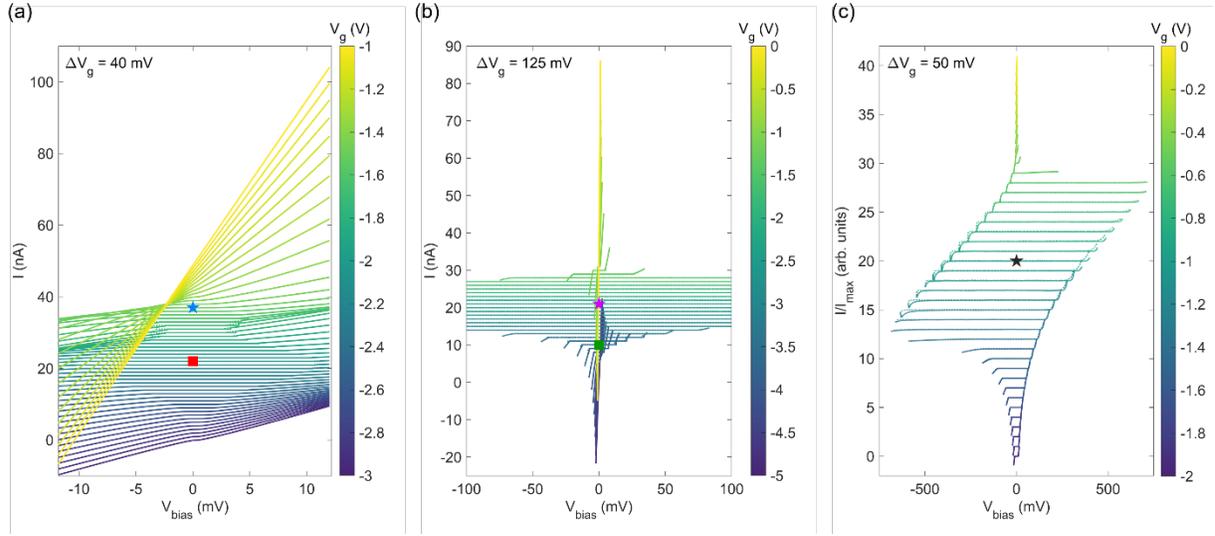

**Supplementary Figure 4**: Raw I-V traces across a range of gate voltages for the (a) 16 nm film and (b) 13 nm film. Stars (squares) mark the traces corresponding to the charge neutrality point (Wigner solid). Traces are offset by 1 nA for clarity. (c) I-V across a larger bias range measured on a different device fabricated from the same 13 nm film. Traces are offset by 1. Solid (dashed) traces correspond to the forward (reverse) bias sweep direction.

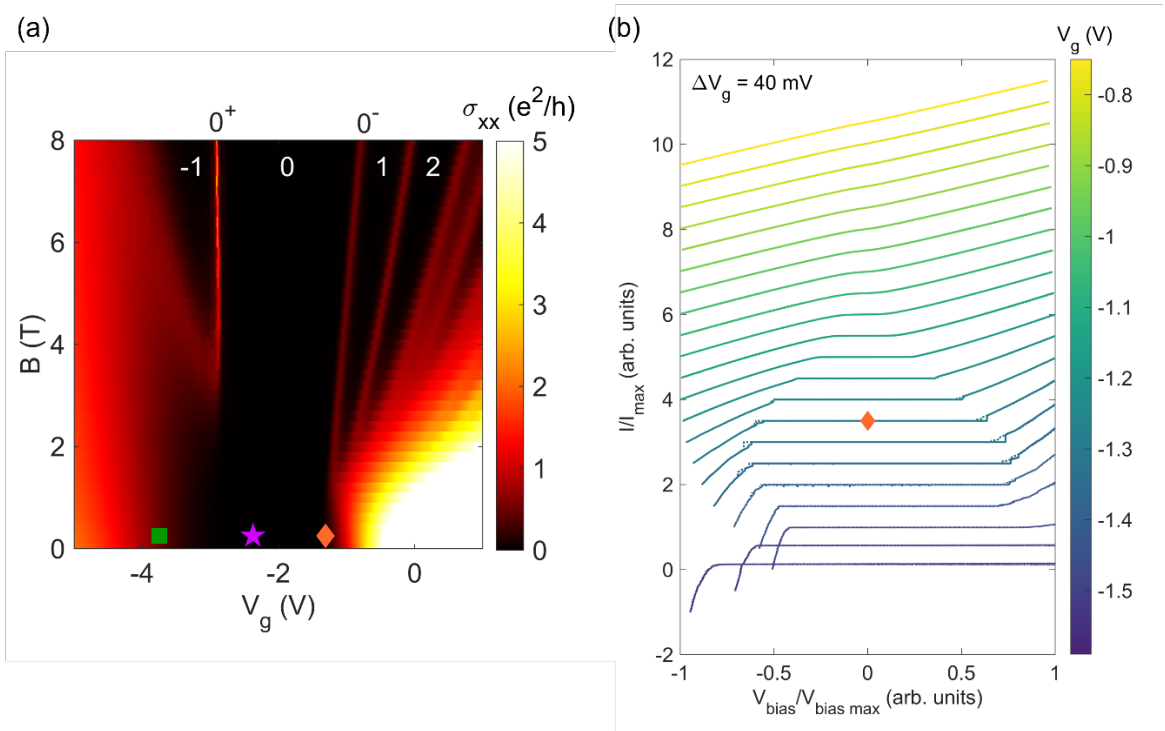

**Supplementary Figure 5**: Electron Wigner solid in the 13 nm film. (a) Landau level spectrum of the 13 nm film at 2 K. The orange diamond marks the electron Wigner solid. (b) I-V traces across a range of gate voltages near the conduction subband edge at 12 mK. Traces are offset by 0.5. Solid (dashed) traces correspond to the forward (reverse) bias sweep direction.

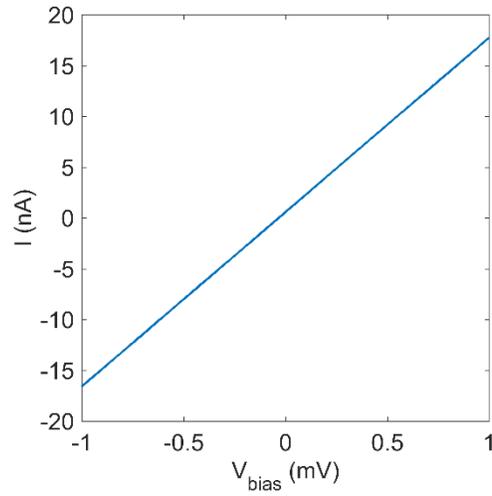

**Supplementary Figure 6**: I-V trace in the $\nu = -1$ quantum Hall plateau at 6 T and $V_g$ = -1.9 V in the 16 nm film.

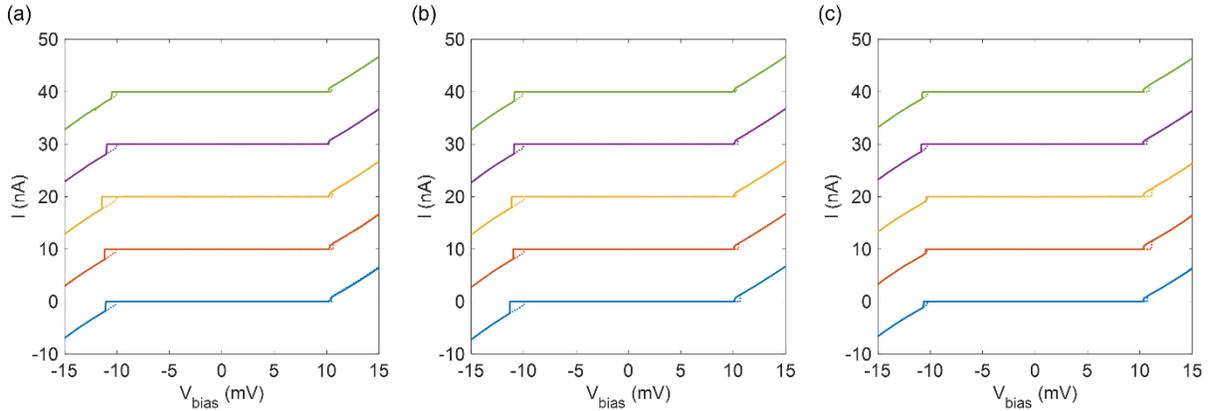

**Supplementary Figure 7**: Bias sweep rate dependence of I-V hysteresis in the Wigner solid phase of the 16 nm film ($V_g$ = -2.08 V, 12 mK). (a) $dV_{bias}/dt = 10^{-5}\ V/s$, (b) $dV_{bias}/dt = 5 \times 10^{-5}\ V/s$, and (c) $dV_{bias}/dt = 10^{-4}\ V/s$. The dashed (solid) traces correspond to the forward (reverse) sweep direction. Traces are offset by 10 nA.

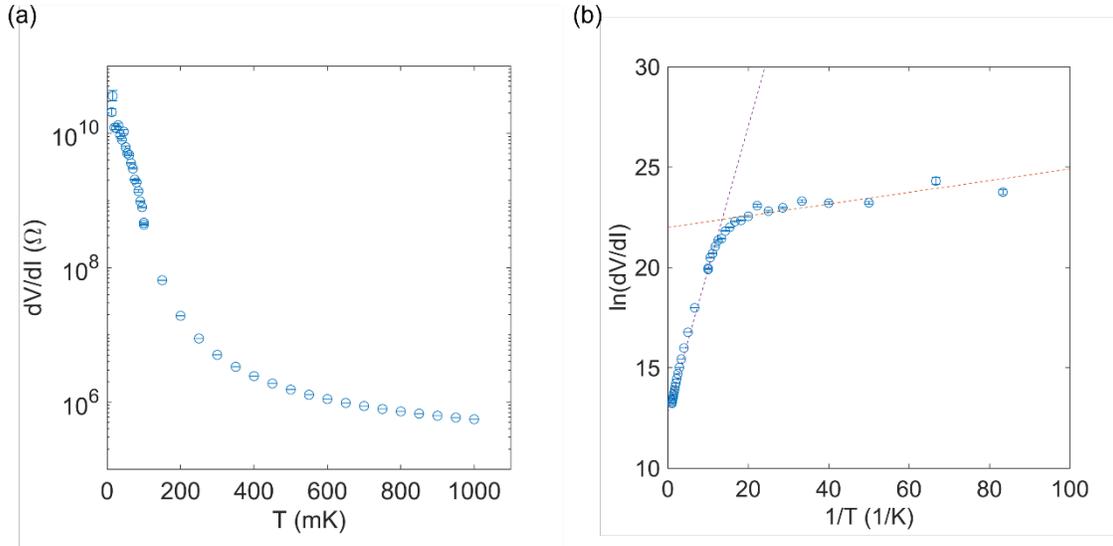

**Supplementary Figure 8**: Temperature-dependent resistance of the Wigner solid in the 16 nm film. (a) Differential resistance at zero bias up to 1 K. (b) Arrhenius plot of differential resistance. Dashed lines indicate two linear fits which cross at ~ 65 mK.

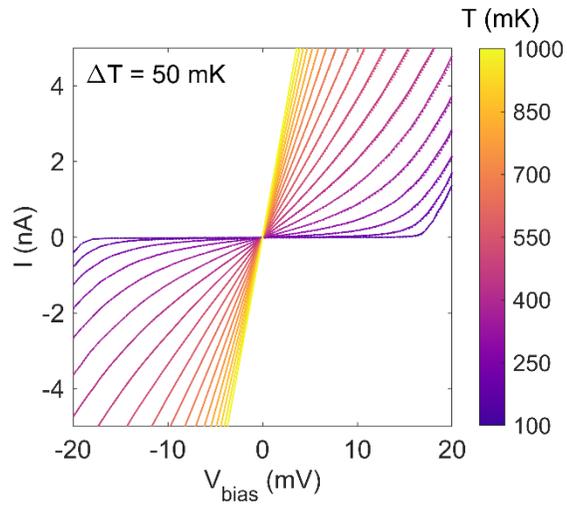

**Supplementary Figure 9:** Temperature-dependence of I-V up to 1 K in the hole-like Wigner solid phase of the 13 nm film ($V_g$ = -3.75 V).

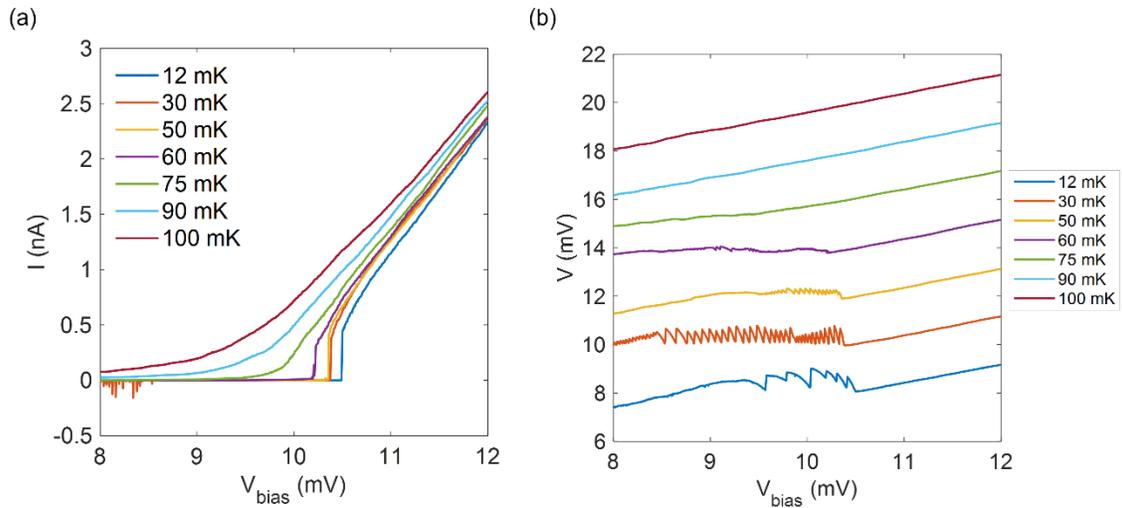

**Supplementary Figure 10**: Temperature-dependence of (a) I-V and (b) voltage fluctuations in the Wigner solid phase of the 16 nm film. Traces are offset by 2 mV in (b).

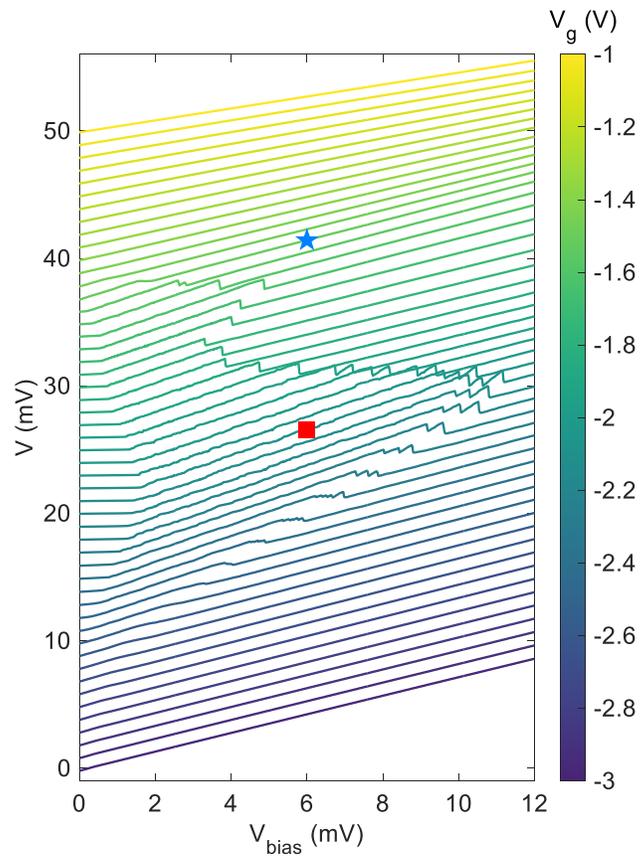

**Supplementary Figure 11**: Gate-dependent voltage fluctuations in the 16 nm film at 12 mK. Star (square) marks the charge neutrality point (Wigner solid). Traces are offset by 1 mV.

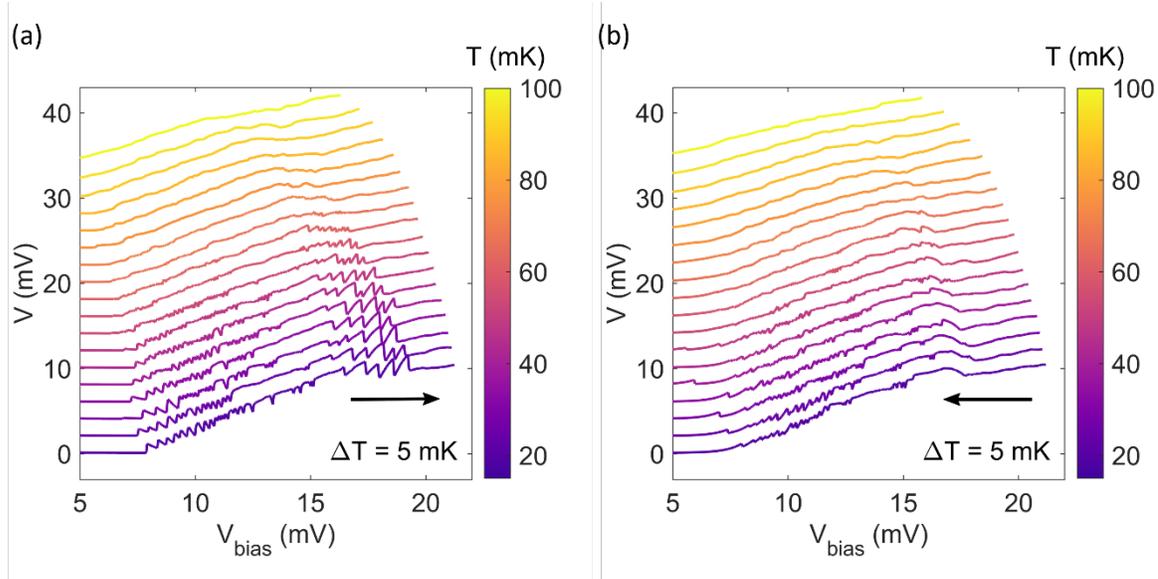

**Supplementary Figure 12**: Voltage fluctuations in the hole-like Wigner solid of the 13 nm film across a range of temperatures in the (a) forward and (b) reverse bias sweep direction. Traces are offset by 2 mV.

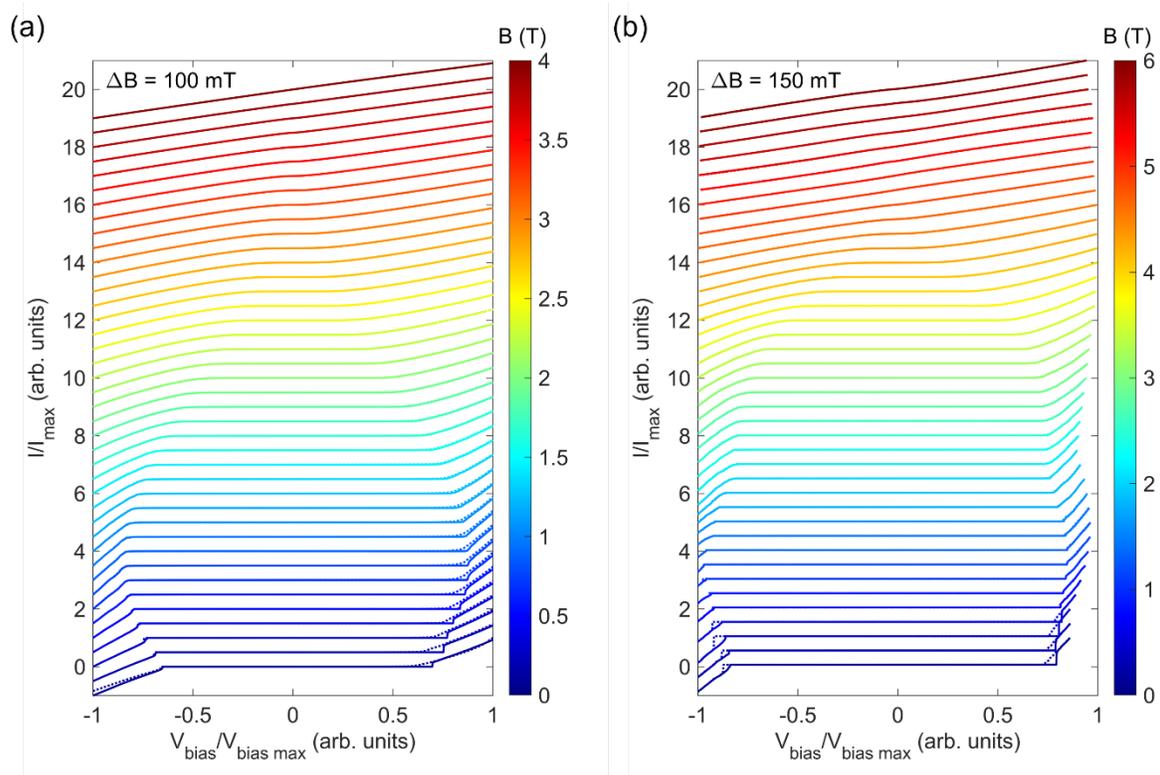

**Supplementary Figure 13**: Magnetic field dependent I-V of (a) 16 nm ($V_g$ = -2.08 V) and (b) 13 nm ($V_g$ = -3.75 V) films in the Wigner solid phase at 12 mK. The magnetic field direction is out-of-plane. Traces are individually normalized in bias voltage and current by their maxima and offset by 0.5 for clarity. Solid (dashed) traces correspond to the forward (reverse) bias sweep direction.

**Supplementary Note 1**

For the Landau level spectrum of the 16 nm film (Fig. 1a), a constant current of 100 pA was supplied at 17.778 Hz with a current source (CS580, Stanford Research Instruments). Longitudinal ($V_{xx}$) and transverse ($V_{xy}$) voltages were measured with lock-in amplifiers (SR860, Stanford Research Instruments) using a time constant of 300 ms with 24 dB of roll-off. A constant current was chosen to avoid current transients as the sample resistance changes, which were found empirically to have a long time constant due to cable capacitance and introduced artifacts into the measurement. A voltage $V_g$ was supplied from a DC voltage source (DC205, Stanford Research Instruments) to the gate to modulate carrier density.

For the Landau level spectrum of the 13 nm film (Fig. 1c), a voltage excitation of 100 mV RMS from the lock-in's sine out biased a 100 MΩ resistor to supply a constant current (Fig. S1b). The voltages and current through the device were measured with the same lock-in settings as mentioned above. Due to the exceedingly large resistance of the charge neutrality gap in this film, a shunt resistor of 100 kΩ was placed in parallel with the sample to avoid overheating the device. This large resistance overloads the cryogenic filters in our dilution refrigerator, which is deleterious to the measurement bandwidth. Therefore, we measured the device at 2 K in a He-4 cryostat with shorter cabling and without cryogenic filters.

Resistances were calculated as $R_{xx,xy} = V_{xx,xy}/I$ and the corresponding conductivities were calculated by tensor inversion using the following equations:

$$\sigma_{xx} = \frac{\rho_{xx}}{\rho_{xx}^2 + \rho_{xy}^2}, \sigma_{xy} = -\frac{\rho_{xy}}{\rho_{xx}^2 + \rho_{xy}^2}.$$

For the DC measurements, a voltage bias was supplied from a DC voltage source while voltage and current preamplifiers paired with voltmeters measured V and I, respectively.

**Supplementary Note 2**

The raw data from the normalized I-V traces in Figs. 3a and 3b in the main text are shown above in Figs. S3a and S3b. The resistance changes dramatically as a function of gate voltage which obscures key qualitative features of the raw data. Therefore, we chose to normalize the traces individually by their maximum applied bias and measured current in the main text. In the 16 nm film, a maximum bias of 12 mV was applied at all gate voltages, which was sufficient to access the linear regime. The 13 nm film is much more resistive, so we programmatically set a power limit of 50 pW and bias limit of 100 mV to avoid catastrophic breakdown of the device.

In the charge neutrality gap of the 13 nm film (Fig. S3b), the I-V is flat, and the current is below the noise threshold up to 100 mV due to its exceedingly high resistance. At very large bias (Fig. S3c), the I-V becomes non-linear in the charge neutrality gap but is distinct from the non-linear I-V seen in the Wigner solid. The current exponentially increases and is strongly hysteretic. We attribute this to breakdown of the gap due to thermionic emission. This explanation is supported by the non-reciprocity in the I-V which evolves with gate voltage. Above the charge neutrality point, the threshold for breakdown is shifted toward positive bias because the Fermi level is near the conduction subband edge. Therefore, a small negative bias lifts the Fermi level and excites electrons over the gap. The opposite case holds for traces taken below the charge neutrality point, where the Fermi level is near the valence subband edge. In this case, a small positive bias pushes the Fermi level down and excites holes over the gap. This smooth exponential I-V is distinct from the sharp double-threshold I-V seen in the Wigner solid and occurs at biases far in excess of the typical depinning threshold of the Wigner solid (~5-35 mV). Additionally, the I-V traces in the Wigner solid phase do not show a significant degree of non-reciprocity.

**Supplementary Note 3**

The contact resistance may become very large at low carrier densities where the Wigner solid appears, which may add spurious non-linearities in the two-point I-V traces. In the following, we use the two-point resistance in a quantum Hall plateau to estimate the contact resistance in the Wigner solid. We tune the gate voltage to a carrier density within the Wigner solid phase and apply an out-of-plane magnetic field to access the $\nu = -1$ quantum Hall plateau (see Fig. 1a). The two-point resistance is estimated from the slope of the I-V to be 58.2 kΩ (Fig S5).

In our device structure, the gated region with filling factor $\nu'$ is flanked on either side by an ungated region with filling $\nu$ which forms a bipolar junction. The theoretical two-point resistance without contact resistance in the quantum Hall regime is [1]:

$$R_{2\text{pt}}(\nu, \nu') = \frac{h}{e^2} \frac{2|\nu'| + |\nu|}{|\nu'\nu|}$$

At 6 T, the ungated region of $Cd_3As_2$ is in the $\nu = 4$ plateau, so $R_{2pt}(4, -1) = (3/2)h/e^2 \approx 38.7$ kΩ. Additionally, the cryogenic filters used in our measurement provide a series resistance of 6 kΩ. Therefore, the contact resistance is ~7 kΩ per contact, which is negligibly small compared to the Wigner solid (> 1 MΩ).

**Supplementary Reference**